# Computer Simulation Codes for the Quine-McCluskey Method of Logic Minimization

Sourangsu Banerji
Department of Electronics & Communication Engineering,
RCC-Institute of Information Technology

## ABSTRACT
The Quine-McCluskey method is useful in minimizing logic expressions for larger number of variables when compared with minimization by Karnaugh Map or Boolean algebra. In this paper, we have tried to put together all of the computer codes which are available on the internet, edited and modified them as well as rewritten some parts of those collected codes our self, which are used in the implementation of the Quine-McCluskey method. A brief introduction and the logic of this method are discussed following which the codes have been provided. The Quine-McCluskey Method has been implemented using computer languages like C and C++ using some amount of variations. Our effort is to list them all, so that the readers well versed in any of the particular computer language will find it easy to follow the code written in that particular language.

## KEYWORDS

## I. INTRODUCTION
The Quine–McCluskey algorithm or the method of prime implicants is a method used for minimization of boolean functions. It was developed by W.V. Quine and Edward J. McCluskey in 1956. It is functionally identical to Karnaugh mapping, but the tabular form makes it more efficient for use in computer algorithms, and it also gives a deterministic way to check that the minimal form of a Boolean function has been reached. It is sometimes referred to as the tabulation method.
The method involves two steps:
1. Finding all prime implicants of the function.
2. Use those prime implicants in a prime implicant chart to find the essential prime implicants of the function, as well as other prime implicants that are necessary to cover the function.
In this paper, we intend to discuss the Quine-McCluskey minimization procedure as well as provide the readers with all the simulation codes which are available on net in one single paper, highlighting the variations in each of the given codes implemented using a different computer language. The procedure which is discussed in the following section 2 and 3 has also been taken from the net and for that appropriate references have been given.

## II. QUINE-McCLUSKEY MINIMIZATION PROCEDURE
This is basically a tabular method of minimization and as much it is suitable for computer applications. The procedure for optimization as follows:

**Step 1:** Describe individual minterms of the given expression by their equivalent binary numbers.
**Step 2:** Form a table by grouping numbers with equivalent number of 1's in them, i.e. first numbers with no 1's, then numbers with one 1, and then numbers with two 1's, … etc.
**Step 3:** Compare each number in the top group with each minterm in the next lower group. If the two numbers are the same in every position but one, place a check sign (✓) to the right of both numbers to show that they have been paired and covered. Then enter the newly formed number in the next column (a new table). The new number is the old numbers but where the literal differ, an "x" is placed in the position of that literal.
**Step 4:** Using (3) above, form a second table and repeat the process again until no further pairing is possible. (On second repeat, compare numbers to numbers in the next group that have the same "x" position.
**Step 5:** Terms which were not covered are the prime implicants and are ORed and ANDed together to form final function.
**Note:** The procedure above gives you the prime implicant but not **essential** prime implicant.

**Example 1**
Minimize the function given below by Quine-McCluskey method.

| f(A,B,C,D)= | $\bar{A}\bar{B}\bar{C}\bar{D}$+ | $\bar{A}B\bar{C}D$+ | $\bar{A}B C\bar{D}$+ | $A\bar{B}\bar{C}D$+ | $A\bar{B}C\bar{D}$+ | $AB\bar{C}D$+ | $ABC\bar{D}$+ | $ABCD$+ | $\bar{A}BCD$ |
|---|---|---|---|---|---|---|---|---|---|
| Binary | 0000 | 0101 | 0110 | 1001 | 1010 | 1101 | 1110 | 1111 | 0111 |
| minterm | 0 | 5 | 6 | 9 | 10 | 13 | 14 | 15 | 7 |
| No of 1's | 0 | 2 | 2 | 2 | 2 | 3 | 3 | 4 | 3 |
| group | 1 | 2 | 2 | 2 | 2 | 3 | 3 | 4 | 3 |

This can be written as a sum of minterms as follows:
$$f(A,B,C,D) = \sum m(0, 5, 6, 7, 9, 10, 13, 14, 15)$$

**Step 1:** Form a table of functions of minterms according to the number of 1's in each minterm as shown in Table E1.a

| minterm | A | B | C | D | | |
|---|---|---|---|---|---|---|
| * 0 | 0 | 0 | 0 | 0 | | } All numbers with no 1's in each minterm (a) |
| 5 | 0° | 1 | 0° | 1 | ✓✓… | |
| 6 | 0 | 1 | 1 | 0 | ✓✓… | } All numbers with two 1's in each minterm |
| 9 | 1 | 0 | 0 | 1 | ✓✓… | |
| 10 | 1 | 0 | 1 | 0 | ✓✓… | |
| 7 | 0 | 1 | 1° | 1 | ✓✓… | |
| 13 | 1° | 1 | 0 | 1 | ✓✓… | } All numbers with three 1's in each minterm |
| 14 | 1 | 1 | 1 | 0 | ✓✓… | |
| 15 | 1 | 1 | 1 | 1 | ✓ | } All numbers with four 1's in each minterm |

| AB\CD | 00 | 01 | 11 | 10 |
|---|---|---|---|---|
| 00 | 1 [0] | [4] | [12] | [8] |
| 01 | [1] | 1 [5] | 1 [13] | 1 [9] |
| 11 | [3] | 1 [7] | 1 [15] | [11] |
| 10 | [2] | 1 [6] | 1 [14] | 1 [10] |

Table E1.a

**Step 2:** Start pairing off each element of first group with the next, however since $m_0$ has no 1's, it and the next group of numbers with one 1's are missing, therefore they cannot be paired off. Start by pairing elements of $m_5$ with $m_7$, $m_{13}$, $m_{14}$, and $m_6$ with $m_7$, $m_{13}$, $m_{14}$, and so on… If they pair off, write them in a separate table and ✓ the minterm that pair, i.e. $m_5$ and $m_7$ pair off 0101 and 0111 to produce 01x1, so in the next table E1.b under "minterm paired" we enter "5, 7" and under "ABCD" we enter "01x1" and place a ✓ sign in front of 5 and 7 in Table E1.a

**Note:** Each minterm in a group must be compared with every minterm in the other group even if either or both of them have already been checked✓.

| minterms paired | A B C D |
|---|---|
| 5, 7 | 0 1 X 1 ✓ |
| 5, 13 | X 1 0 1 ✓✓ |
| 6, 7 | 0 1 1 X ✓✓ |
| 6, 14 | X 1 1 0 ✓✓ |
| 9, 13 | 1 X 0 1  ………… (b) |
| 10, 14 | 1 X 1 0  ………… (c) |
| 7, 15 | X 1 1 1 ✓✓ |
| 13, 15 | 1 1 X 1 ✓✓ |
| 14, 15 | 1 1 1 X ✓✓ |

Table E1.b

| Paired minterms from E1.b | A | B | C | D | |
|---|---|---|---|---|---|
| 5,7 – 13,15 | x | 1 | x | 1 | … (d) |
| 6,7 – 14,15 | x | 1 | 1 | x | … (e) |

Table E1.c

**Step 3:** Now repeat the same procedure by pairing each element of a group with the elements of the next group for elements that have "x" in the same position. For example, "5,7" matches "13,15" to produce x1x1. These elements are placed in table E1.c as shown, and the above elements in Table E1.b are ✓ checked. (The elements that produce the same ABCD pattern are eliminated.) Since 9,13 and 10,14 in Table E1.b do not pair off, they are prime implicants and with $m_0$, from E1.a, and (d) and (e) from E1.c are unpaired individuals. Therefore, it is possible to write the minimized SOP as a+b+c+d+r or

$$f(A, B, C, D) = \overline{A}\overline{B}\overline{C}\overline{D} + A\overline{C}D + AC\overline{D} + BD + BC$$

**Note:** Check this result for Example 1 by Karnaugh map approach.

Two-square implicants:

| AB\CD | 00 | 01 | 11 | 10 |
|---|---|---|---|---|
| 00 | 1 ⁰ |  ⁴ |  ¹² |  ⁸ |
| 01 |  ¹ | 1 ⁵ | 1 ¹³ | 1 ⁹ |
| 11 |  ³ | 1 ⁷ | 1 ¹⁵ |  ¹¹ |
| 10 |  ² | 1 ⁶ | 1 ¹⁴ | 1 ¹⁰ |

Table E1.b represents all possible two-square implicants and the literals that they eliminate, i.e. 9 ($1001_b$) combined with 13 ($1101_b$) produces 1x01. As a result, literal "B" is eliminated. Corresponding product is $A\overline{C}D$. Since the only way of making an implicant that contains $m_9$ is to combine it with $m_{13}$, the implicant 9-13 is a prime one. The same rule applies to $m_{10}$.

Four-square implicants:

| AB\CD | 00 | 01 | 11 | 10 |
|---|---|---|---|---|
| 00 | 1  (0) |  (4) |  (12) |  (8) |
| 01 |  (1) | 1 (5) | 1 (13) | 1 (9) |
| 11 |  (3) | 1 (7) | 1 (15) |  (11) |
| 10 |  (2) | 1 (6) | 1 (14) | 1 (10) |

Table E1.c represents all possible four-square implicants and the literals that they eliminate, i.e. 5 ($0101_b$) combined with 7 ($0111_b$) and 13 ($1101_b$) and 15 ($1111_b$) produces x1x1. As a result, literals "A" and "C" are eliminated. Corresponding product is BD.

## III. QUINE-McCLUSKEY MINIMIZATION PROCEDURE (Decimal Notation)

**Step 1:** List the minterms grouped according to the number of 1's in their binary representation in the decimal format.

**Step 2:** Compare each minterm with larger minterms in the next group down. If they differ by a power of 2 then they pair-off. Check both minterms and form a second table by the minterms paired and substitute the decimal difference of the corresponding minterms in the bracket, i.e. $m_x$, $m_y$ (y-x).

**Step 3:** Compare each element of the group in the new table with elements of the next lower group and select numbers that have the same numbers in parenthesis. If the lowest minterm number of the table formed in the lower group is greater than the corresponding number by a power of 2 then they combine; place a ✓ on the right of both elements.

**Step 4:** Form a second table by all four minterms followed by both powers of 2 in parentheses, i.e. the previous value (the difference) and the power of 2 that is greater.

**Step 5:** Select the common literals from each prime implicant by comparison.

**Step 6:** Write the minimal SOP from the prime implicant that are not checked✓.

**Note:** Read the above procedure in conjunction with the worked example given below.

## Example E1.1

Minimize the function f given below by Quine-McCluskey method using decimal notation.

$$f(A,B,C,D) = \overline{A}\overline{B}\overline{C}\overline{D} + \overline{A}B\overline{C}D + \overline{A}BC\overline{D} + \overline{A}BCD + A\overline{B}\overline{C}D + A\overline{B}C\overline{D} + AB\overline{C}D + ABC\overline{D} + ABCD$$

**Solution**

**Step 1:** Organize minterm as follows:

$$f(A,B,C,D) = \sum m(0,5,6,7,9,10,13,14,15)$$

Arrange minterms to correspond to their number of 1's as shown in E1.1a

|   | 1's | Minterms |     |
|---|-----|----------|-----|
| * | 0   | 0        | …(a)|
|   |     | 5 ✓      |     |
|   |     | 6 ✓      |     |
|   | 2   | 9 ✓      |     |
|   |     | 10 ✓     |     |
|   |     | 7 ✓      |     |
|   | 3   | 13 ✓     |     |
|   |     | 14 ✓     |     |
|   | 4   | 15 ✓     |     |

Table E1.1a

|   | minterm | paired |     |       |
|---|---------|--------|-----|-------|
|   | 5$^\Phi$, 7$^\Phi$ | (2)$^\dagger$ ✓ |       |
|   | 5, 13   | (8)    | ✓   |       |
|   | 6, 7    | (1)    | ✓   |       |
|   | 6, 14   | (8)    | ✓   |       |
| * | 9, 13   | (4)    |     | …(b)  |
| * | 10, 14  | (4)    |     | …(c)  |
|   | 7, 15   | (8)    | ✓   |       |
|   | 13, 15  | (2)    | ✓   |       |
|   | 14, 15  | (1)    | ✓   |       |

Table E1.1b

|   | minterms paired |        |       |
|---|-----------------|--------|-------|
| * | 5,7-13,15$^i$   | (2,8)  | …(d)  |
| * | 6,14-7,15       | (1,8)  | …(e)  |

Table E1.1c

Φ - Squares combined (2 squares);

† - Number in bracket shows the literal being eliminated, i.e. (2) represents C [A=8, B=4, C=2, D=1];

¡- squares combined (4 squares) and numbers in the brackets are the literals eliminated.

**Step 2:** Compare each element of a group with the element of the next group if the difference is a power of 2 then they pair off, i.e. the first element in group 2 is paired say with the first element in group 3, which is 7-5=2, which is power of 2. Therefore, pair (5,7) makes the first element of the next table and minterms 5 and 7 get checked ✓. The result is shown in Table E1.1b.

**Step 3:** Now for the $2^3$-table again compare each element of the group with elements of the lower group that have the same number in parentheses. If the lowest minterm in the lower group was greater by a power of 2 then they combine, i.e. 5,7 and 13,15 are combined because they have (2) in parentheses and 13 is greater then 5 by 8. Then they are paired off and entered in the next table E1.1c with the original (2) and their difference (8) in the parentheses.

**Step 4:** What we are left with is (a) from Table E1.1a, (b) and (c) from Table E1.1b, and (d) and (e) from Table E1.1c. From Table E1.1c, "d" is 5,7-13,15 (2,8). That means that positions $2^1$ and $2^3$ are X's. Thus, "d" represents function BD. From the same table, "e" is 6,14-7,15 (1,8), which means positions $2^0$ and $2^3$ are X's.

Thus, "e" represents function BC. This can also be obtained by writing the elements of minterms and selecting two remaining literals:

$$\begin{cases} \begin{array}{c|cccc} 6 & 0 & 1 & 1 & 0 \\ 14 & 1 & 1 & 1 & 0 \\ 7 & 0 & 1 & 1 & 1 \\ 15 & 1 & 1 & 1 & 1 \\ \hline & x & 1 & 1 & x \\ & & B & C & \end{array} \end{cases}$$

Therefore, the minimized SOP is

$$f = a + b + c + d + e = \overline{A}\overline{B}\overline{C}\overline{D} + A\overline{C}D + AC\overline{D} + BD + BC$$

**Note:** Compare this with the method of K-map or standard Quine-McCluskey (the first approach).

The above function consists of prime implicants. However, not all of them are necessary essential prime implicants.

**Example 1.1.1. Determination of Essential Prime Implicants**
For the SOP obtained in Example 1.1, determine the essential prime implicants and see if further reduction is possible.

**Solution:**
Construct a prime implicants table as shown in Table 1.1.1a, with prime implicants on left and minterms on top:

|  | Minterms / Prime implicants | 0 | 5 | 6 | 7 | 9 | 10 | 13 | 14 | 15 |
|---|---|---|---|---|---|---|---|---|---|---|
| *(2) | 5, 7 – 13, 15 |  | ✓ |  | ✓ |  |  | ✓ |  | ✓ |
| *(3) | 6, 14 – 7, 15 |  |  | ✓ | ✓ |  |  |  | ✓ | ✓ |
| *(4) | 9, 13 |  |  |  |  | ✓ |  | ✓ |  |  |
| *(5) | 10, 14 |  |  |  |  |  | ✓ |  | ✓ |  |
| *(1) | 0 | ✓ |  |  |  |  |  |  |  |  |
|  |  | ✓(1) | ✓(2) | ✓(3) |  | ✓(4) | ✓(5) |  |  |  |
|  | 5 |  | ✓ |  | ✓ |  |  | ✓ |  | ✓ |
|  | 6 |  |  | ✓ |  |  |  |  | ✓ |  |
|  | 9 |  |  |  |  | ✓ |  |  |  |  |

**Table 1.1.1a**

In each row, (except the bottom) checks ✓ are placed in the columns corresponding to minterms contained in the prime implicant listing in they row, i.e. the first prime implicant testing contains 5, 7, 13, 15. So, ✓ is placed in the first row in columns 5, 7, 13, 15. Repeat for each prime implicant.

Now inspect the table for columns that contain only one✓. That means that that prime implicant is the only term that contains that minterm, i.e. for example $m_0$ must be included in the SOP. This is marked with asterisks (*) in the left column and place ✓ in the bottom row. The same applies to 6, 9, and 10. Therefore, all prime implicants in this example are essential prime implicants. Other empty cells in the bottom row are covered by essential prime implicants. For example, once 5 are selected, and then 7, 13, and 15 also can be ✓ from the bottom row, and so on.

## IV. COMPUTER SIMULATION CODES

## Code for Quine-McCluskey Method in C

### Variant-I

```c
#include <stdio.h>
#include <stdlib.h>
#include <string.h>
#include <math.h>
#include <limits.h>
#include <assert.h>
#include <stdbool.h>
#include <signal.h>
#include <inttypes.h>
#include <float.h>
#include <ctype.h>
#include <time.h>

int *in,**d1,**d2,x,y,**g,**d;
void create(int x,int y);
int staging(int x,int y);
int duplication(int x,int y);
int indexing(int x,int y3,int y);
void pimp(int x,int y3,int a);
void decode(int x,int y3);
void wxyz(int x,int y3);

int main()
{
  int i,j,y1,y2,y3,a,q;
  printf("\n Please give the number of variables you want to minimize - ");
  scanf("%d",&x);
  printf("\n\n In this program your inputs are designated as : ");
  for(j=x-1;j>=0;j--)
  printf("a[%d]",j);
  printf("\n\n Please give the number of minterms that you want to minimize - ");
  scanf("%d",&y);
  in=(int *)malloc(y * sizeof(int));
  d=d1=(int **)malloc(y * sizeof(int *));
  for(i=0;i<y;i++)
  d[i]=d1[i]=(int *)malloc((x+1)*sizeof(int));
```

```c
  for(i=0;i<y;i++)
    {
    printf("\n Please give decimal indices of minterms one at a time :: ");
    scanf("%d",&in[i]);}
    create(x,y);
    y1=y*(y+1)/2;
    d2=(int **)malloc(y1*sizeof(int *));
    for(i=0;i<y1;i++)
    d2[i]=(int *)malloc((x+1)*sizeof(int));
    y2=staging(x,y);
    y3=duplication(x,y2);
    a=indexing(x,y3,y);
    pimp(x,y3,a);
    printf("\n\nThe essential prime implicants giving minimized expression are:\n\n");
    decode(x,y3);
    }
  void create(int x,int y)
    {
    int i,j,a;
    for(i=0;i<y;i++)
      {
      a=in[i];
      for(j=0;j<x;j++)
        {
        d[i][j]=d1[i][j]=a%2;
        a=a/2;
        }
      d[i][x]=d1[i][x]=8;
      }
    }
  int staging(int x,int y)
    {
    int i1,j1,k1,t1,i2,j2,t2,c;
    i2=0;c=0;
    for(i1=0;i1<(y-1);i1++)
      {
      for(j1=i1+1;j1<y;j1++)
        {
        t1=0;
        for(k1=0;k1<x;k1++)
          {
          if(d1[i1][k1]!=d1[j1][k1])
            {
            t1++;
            t2=k1;
            }
```

```
      }
    if(t1==1)
      {
      for(j2=0;j2<t2;j2++)
      d2[i2][j2]=d1[i1][j2];
      d2[i2][t2]=3;
      for(j2=t2+1;j2<y;j2++)
      d2[i2][j2]=d1[i1][j2];
      d2[i2][x]=8;
      d1[i1][x]=9;
      d1[j1][x]=9;
      i2++;
      }
      }
    }
  for(i1=0;i1<y;i1++)
    {
    if(d1[i1][x]==8)
      {
      for(j1=0;j1<=x;j1++)
      d2[i2][j1]=d1[i1][j1];
      i2++;
      }
    }
  for(j1=0;j1<x;j1++)
    {
    if(d1[0][j1]==d2[0][j1])
    c++;
    }
  if(c<x)
    {
    d1=(int **)malloc(i2*sizeof(int *));
    for(i1=0;i1<i2;i1++)
    d1[i1]=(int *)malloc((x+1)*sizeof(int));
    for(i1=0;i1<i2;i1++)
      {
      for(j1=0;j1<=x;j1++)
      d1[i1][j1]=d2[i1][j1];
      }
    staging(x,i2);
    }
  else
  return(i2);
  }
int duplication(int x,int y)
  {
```

```c
    int i1,i2,j,c,t;
    t=0;
    for(i1=0;i1<(y-1);i1++)
      {
      for(i2=i1+1;i2<y;i2++)
         {
         c=0;
         for(j=0;j<x;j++)
            {
            if(d1[i1][j]==d1[i2][j])
            c++;
            }
         if(c==x)
         d1[i2][x]=9;
         }
      }
    for(i1=0;i1<y;i1++)
        {
        if(d1[i1][x]==9)
        t++;
        }
      i2=y-t;
      d2=(int **)malloc(i2*sizeof(int *));
      for(j=0;j<i2;j++)
      d2[j]=(int *)malloc((x+1)*sizeof(int));
      i2=0;
      for(i1=0;i1<y;i1++)
      if(d1[i1][x]==8)
         {
         for(j=0;j<=x;j++)
         d2[i2][j]=d1[i1][j];
         i2++;
         }
    return(i2);
    }
  int indexing(int x,int y3,int y)
    {
    int i1,j,c1,i2,c,a;
    c=0;a=1;
    for(j=0;j<x;j++)
    if(d1[0][j]==3)c++;
    for(i1=0;i1<c;i1++)
    a=a*2;
    g=(int **)malloc(y3*sizeof(int *));
    for(j=0;j<y3;j++)
    g[j]=(int *)malloc(a*sizeof(int));
```

```
   for(i1=0;i1<y3;i1++)
   for(j=0;j<a;j++)
   g[i1][j]=-2;
   for(i1=0;i1<y3;i1++)
     {
      c=0;
      for(i2=0;i2<y;i2++)
        {
         c1=0;
         for(j=0;j<x;j++)
           {
            if((d2[i1][j]==d[i2][j])||(d2[i1][j]==3))
            c1++;
            if(c1==x)
              {
               g[i1][c]=in[i2];
               c++;
              }
           }
        }
     }
   return(a);
  }
 void pimp(int x,int y3,int a)
   {
   int i1,i2,j1,j2,c,w,j3,c1,c2,j4,c3,c4;
   c=0;
   for(i1=0;i1<y3;i1++)
     {
      for(j1=0;j1<a;j1++)
        {
         if(g[i1][j1]!=-2)
           {
            for(i2=0;i2<y3;i2++)
              {
               if(i2!=i1)
                 {
                  for(j2=0;j2<a;j2++)
                  if(g[i1][j1]!=g[i2][j2])
                  c++;
                 }
              }
            if(c==a*(y3-1))
            d2[i1][x]=91;c=0;
           }
        }
```

```
    }
 for(i1=0;i1<y3;i1++)
   {
    if(d2[i1][x]==91)
      {
       for(j1=0;j1<a;j1++)
         {
          if(g[i1][j1]!=-2)
            {
             for(i2=0;i2<y3;i2++)
               {
                if(i1!=i2)
                  {
                   for(j2=0;j2<a;j2++)
                   if(g[i1][j1]==g[i2][j2])
                   g[i2][j2]=-3;
                  }
               }
            }
         }
      }
   }
 for(i1=0;i1<y3;i1++)
   {
    if(d2[i1][x]==91)
      {
       for(j1=0;j1<a;j1++)
       if(g[i1][j1]!=-2)g[i1][j1]=-1;
      }
   }
 for(i1=0;i1<y3;i1++)
   {
    if(d2[i1][x]!=91)
      {
       for(j1=0;j1<a;j1++)
         {
          if(g[i1][j1]>=0)
            {
             for(i2=0;i2<y3;i2++)
             if(i2!=i1)
               {
                for(j2=0;j2<a;j2++)
                  {
                   if(g[i2][j2]>=0)
                     {
                      if(g[i1][j1]==g[i2][j2])
```

```
                       {
                       w=i2;
                       if((d2[w][x]==90)||(d2[w][x]==8))
                         {
                          for(j3=0,c1=0;j3<x;j3++)
                          if(d2[i1][j3]==3)
                          c1++;
                          for(j3=0,c2=0;j3<x;j3++)
                          if(d2[i2][j3]==3)
                          c2++;
                          if(c1>c2)
                             {
                             d2[i1][x]=90;
                             g[i2][j2]=-1;
                             }
                           if(c2>c1)
                              {
                              d2[i1][x]=8;
                              d2[i2][x]=90;
                              g[i1][j1]=-1;
                              }
                            if(c2==c1)
                              {
                              for(j3=0,c3=0,c4=0;j3<a;j3++)
                                {
                                if(g[i1][j3]==-
                                1)c3++;
                                if(g[i2][j3]==-1)c4++;
                                }
                              if(c3>c4)
                                {
                                d2[i2][x]=90;d1[i1][x]=8;
                                g[i1][j1]=-1;
                                }
                              if(c3==c4)
                                {
                                d2[i1][x]=90;
                                g[i2][j2]=-1;
                                }
                              if(c3<c4)
                                {
                                d2[i1][x]=90;
                                g[i2][j2]=-1;
                                }
                            }
                       }
```

```
                    if(d2[w][x]==91)
                    d1[w][x]=8;
                    }
                  }
                }
              }
            }
          }
        }
      }
  return;
  }
void decode(int x,int y3)
  {
  int i,j;
  for(i=0;i<y3;i++)
    {
    if((d2[i][x]==91)||(d2[i][x]==90))
      {
      for(j=x-1;j>=0;j--)
        {
        if(d2[i][j]==0)printf("a[%d]'",j);
        if(d2[i][j]==1)printf("a[%d]",j);
        }
      }
    printf("\n\n");
    }
  return;
  }
```

The following code does not take into account don't care conditions.

### Variant-II

```
#include <stdio.h>
#include <stdlib.h>
#include <string.h>
#include <math.h>
#include <limits.h>
#include <assert.h>
#include <stdbool.h>
#include <signal.h>
#include <inttypes.h>
#include <float.h>
#include <ctype.h>
#include <time.h>
```

```
/*      Global declarations      */
#define NMAX        12      /* max number of variables */
#define NTERMS      4096    /* 2 to the NMAX           */
#define QTERMS      256     /* NTERMS divided by WORD  */
#define MTERMS      4096    /* maximum numbers of minterms */
                            /* allowed                 */
#define MIMPS       48      /* largest allowable number of */
                            /* prime implicants        */
#define LIN         120     /* longest input line size */
#define BIG         32000   /* short integer infinity  */
#define WORD        16      /* short integer size      */
#define gbit(a,b)   (((a)>>(b))&1)
int nvars;          /* number of variables in case     */
int nterms = 1;     /* numbers of terms in case        */
int nwords;         /* nterms/WORD + 1                 */
int minterm[QTERMS];    /* minterm array (Karnaugh map) */
int noterm[QTERMS];     /* no cares in Karnaugh map    */
int impchart[QTERMS][MIMPS];    /* prime implicant chart */
int impcnt[QTERMS];     /* indicates coverage of a minterm */
int impext[QTERMS];     /* indicates multiple coverage of term */
int essprm[MIMPS];      /* marker for essential prime implicants*/
int imps = 0;           /* number of prime implicants  */
int priterm[MIMPS];     /* term values for prime implicants */
int pricare[MIMPS];     /* nocare values for prime implicants */
int pptr;           /* number of actual nterms         */
struct list {
        int term;
        int mom;        /* first ancestor */
        int dad;        /* second ancestor */
        int nocare;
        int match;
        } plist[MTERMS*2];
char func = 'f';    /* function name */
char vname[NMAX];       /* variable name list */
char *fgets();
main()
{
int i, j;
/* input the functions to be reduced */
getterms();

if ( (nvars > 0 ) && (nvars <= NMAX) ) {
   /* generate Quine-McCluskey reduction */
   quine();
   /* set up prime implicant chart and determine essential primes */
   i = pchart();
```

```
      /* determine minimum function */
      reduction(i);
   }
   getchar();
}
getterms()
{
char(inline[LIN]);
register i=0, j, temp;
int format;
int all = 1, none = 0;        /* degenerate function checks */
while ( i < QTERMS ) minterm[i++] = 0;
/* prompt for format of input */
fprintf(stdout,"Welcome to the Dallen-Rogers switching function\n");
fprintf(stdout,"minimization program, Version 1.0, Dec 1, 1981.\n");
fprintf(stdout,"Enter the number for your preferred input type.\n");
fprintf(stdout,"     1 - truth table\n");
fprintf(stdout,"     2 - decimal codes for Karnaugh mapping\n");
fprintf(stdout,"     3 - logical expression\n");
fprintf(stdout,"Format number? ");
fgets(inline,LIN,stdin);
sscanf(inline,"%d",&format);
if ( (format < 1) || (format > 3) ) {
   fprintf(stdout,"I don't understand your selection. Please try again.\n");
   fprintf(stdout,"format number (1 to 3) ? ");
   fgets(inline,LIN,stdin);
   sscanf(inline,"%d",format);
   if ( (format < 1) || (format > 3) ) {
          fprintf(stderr,"invalid input\n");
          nvars = 0;
          return;
   }
}
switch (format) {
       case 1:
                 ttable();
                 break;
       case 2:
                 kmap();
                 break;
       case 3:
                 expres();
                 break;
}
/* check for degenerate cases */
temp = WORD;
```

```
     if ( nterms < WORD ) temp = nterms;
     for ( i=0; i<nwords; i++) {
        none |= minterm[i];
        for ( j=0; j<temp; j++ ) {
              all &= gbit(minterm[i]|noterm[i],j);
        }
     }
     if ( all != 0 ) {
        fprintf(stdout,"\nfunction = 1\n");
        nvars = 0;
     } else if ( none == 0 ) {
        fprintf(stdout,"\nfunction = 0\n");
        nvars = 0;
     }
     return;
}
ttable()
{
char inline[LIN];
char entry[WORD];
register i=0, j, k;
char *gptr;
/* instructions for inputting the truth table */
fprintf(stdout,"Enter each row of your truth table, with input\n");
fprintf(stdout,"values as 0, 1 or X (not-cares) plus an output\n");
fprintf(stdout,"value of 0 or 1.  Enter an extra RETURN after the\n");
fprintf(stdout,"last line of the truth table. Delimiters between\n");
fprintf(stdout,"column entries are not required\n");
fprintf(stdout,"Row 1: ");
fgets(inline,LIN,stdin);
/*       Read first line of input to determine the number         */
/*       of variables involved                                    */
nvars = 0;
i = 0;
while ( inline[i] != '\n') {
   if ( (inline[i]=='1')||(inline[i]=='0')||(inline[i]=='x')||
        (inline[i]=='X') ) {
           entry[nvars++] = inline[i];
   }
   i++;
}
nvars--;
/*       Determine the maximum numbers of minterms possible  */
/*       and the amount of storage required                  */
for ( i=0; i<nvars; i++ ) {
   nterms *= 2;
```

```
}
nwords = ((nterms+(WORD-1))/WORD);
/* set up default variable names */
for ( i=nvars-1; i>=0; i-- ) {
#ifdef JMH
   vname[i] = (char)('A' + nvars - i - 1);
#else
   vname[i] = (char)('z' - nvars + i + 1);
#endif
}
/* set up defaults as not-cares */
i = 0;
while ( i < nwords ) {
   for ( j=0; j<MIMPS; j++ ) impchart[i][j] = 0;
   noterm[i] = ~0;
   minterm[i++] = 0;
}
/* zero out unused bits */
if (nterms < WORD ) {
   for ( j=nterms; j<WORD; j++ ) {
        noterm[0] &= ~(1<<j);
   }
}
/*      Process each line of truth table and input the next line */
k = 2;
while ( (gptr != NULL) && (inline[0] != '\n') ) {
   i = 0;
   j = 0;
   while ( inline[i] != '\n') {
      if ( (inline[i]=='1')||(inline[i]=='0')||(inline[i]=='x')||
           (inline[i]=='X') ) {
              entry[j++] = inline[i];
      }
      i++;
   }
   ntrmv(0,entry,0);
   fprintf(stdout,"Row %d: ",k++);
   gptr = fgets(inline,LIN,stdin);
}
return;
}
ntrmv(i,entry,trm)
register i;/* column number from input table */
char *entry;       /* pointer to next input term in input string */
register trm;      /* current minterm configuration  */
{
```

```c
   int num;
   char iterm;
   register j;
   /* have all the input terms been processed ? */
   if ( i >= nvars ) {
      sscanf(entry,"%d",&num);
      minterm[trm/WORD] |= num<<(trm%WORD);
      noterm[trm/WORD] &= ~(1<<(trm%WORD));
   } else {
      sscanf(entry,"%1s",&iterm);
      if ( (iterm == '1') || (iterm == '0') ) {
            trm = (trm<<1)|(iterm-'0');
            ntrmv(i+1,entry+1,trm);
      } else {
            /* any printed symbols other than 0 or 1 are treated */
            /* as not-care terms.                                */
            trm = trm<<1;
            ntrmv(i+1,entry+1,trm);
            trm |= 1;
            ntrmv(i+1,entry+1,trm);
      }
   }
   return;
}
kmap()
{
char inline[LIN];
char entry[WORD];
register i=0, j, temp;
int num;
char *gptr;
fprintf(stdout,"Input the number of variables in your function: ");
fgets(inline,LIN,stdin);
sscanf(inline,"%d",&nvars);
/* Check for valid number of variables */
if ( nvars <= 0 ) {
   nvars = 0;
   return;
}
if ( nvars > NMAX ) {
   fprintf(stdout,"Too many variables, %d max\n",NMAX);
   nvars = 0;
   return;
}
/* Determine maximum number of minterms and required storage */
for ( i=0; i<nvars; i++ ) {
```

```
      nterms *= 2;
   }
   nwords = ((nterms+(WORD-1))/WORD);
   for ( i=0; i<nwords; i++ ) {
      for ( j=0; j<MIMPS; j++ ) impchart[i][j] = 0;
   }
   /* set up default variable names */
   for ( i=nvars-1; i>=0; i-- ) {
#ifdef JMH
      vname[i] = (char)('A' + nvars - i - 1);
#else
      vname[i] = (char)('z' - nvars + i + 1);
#endif
   }
   fprintf(stdout,"On one line, input each minterm number ");
   fprintf(stdout,"(0 to %d) separated by spaces or tabs:\n",nterms-1);
   gptr = fgets(inline,LIN,stdin);
   if ( (gptr != NULL) && (inline[0] == '\n') ) {
      nvars = 0;
      return;
   }
   /* Process minterms */
   i = 0;
   while (sscanf(&inline[i],"%s",entry) != EOF  ) {
      j = 0;
      while ( entry[j++] != NULL );
      i += j;
      sscanf(entry,"%d", &num );
      minterm[num/WORD] |= (1<<(num%WORD));
   }
   /* Process not care terms */
   fprintf(stdout,"Input each not-care number (0 to %d), if any:\n",nterms-1);
   gptr = fgets(inline,LIN,stdin);
   if ( (gptr != NULL) && (inline[0] != '\n') ) {
      i = 0;
      while (sscanf(&inline[i],"%s",entry) != EOF  ) {
         j = 0;
         while ( entry[j++] != NULL );
         i += j;
         sscanf(entry,"%d", &num );
         noterm[num/WORD] |= (1<<(num%WORD));
      }
   }
   return;
}
expres()
```

```c
{
char inline[LIN];
char outline[LIN];
char tterm[WORD];
char vdig;
register i=0, j=0, k=0;
int flag = 0;
char *gptr;
fprintf(stdout,"Input a logical expression, in sum of products form,\n");
fprintf(stdout,"using single upper or lower case letters as\n");
fprintf(stdout,"variable names (no subscripts) and a ' following the\n");
fprintf(stdout,"variable name to indicate a complement.  Use a + between\n");
fprintf(stdout,"terms. Spaces, extraneous symbols, and anything\n");
fprintf(stdout,"in front of an optional equal sign are ignored.\n? ");
gptr = fgets(inline,LIN,stdin);
/* check for bad input */
nvars = 0;
if ((gptr==NULL) || (inline[0]=='\n')) {
   nvars == 0;
   fprintf(stderr,"no input expression, program terminated\n");
   return;
}
/* process expression: eliminate white space, count variables, */
/* and reverse position of complement sign.                    */
while (inline[i] != '\n') {
   if ( inline[i++] == '=' ) flag = i;
}
i = flag;
while ( inline[i] != '\n' ) {
   if ( inline[i] == '+' ) {
           outline[j++] = inline[i];
   } else if ( inline[i] == '\'' ) {
           /* reverse complement sign position */
           outline[j] = outline[j-1];
           outline[j-1] = inline[i];
           j++;
   } else if (((inline[i]>='a')&&(inline[i]<='z')) ||
           ((inline[i]>='A')&&(inline[i]<='Z'))) {
           /* check to see if a new variable */
           outline[j++] = inline[i];
           flag = 0;
           for ( k=0; k<nvars; k++ ) {
              if ( vname[k] == inline[i] ) flag = 1;
           }
           if ( flag == 0 ) {
              vname[nvars++] = inline[i];
```

```
            }
        }
        i++;
    }
    /* j is length of outline */
    outline[j++] = '+';
    /* Determine maximum number of minterms and required storage */
    for ( i=0; i<nvars; i++ ) {
        nterms *= 2;
    }
    nwords = ((nterms+(WORD-1))/WORD);
    /* set up defaults as zero terms */
    i = 0;
    while ( i < nwords ) {
        for ( k=0; k<MIMPS; k++ ) impchart[i][k] = 0;
        noterm[i] = 0;
        minterm[i++] = 0;
    }
    /* check for valid expressions */
    if ( nvars == 0 ) {
        fprintf(stderr,"Expression not in readable form\n");
        return;
    }
    /* Convert variables to minterms */
    k = 0;
    while ( k < j ) {
        tterm[nvars] = '1';         /* Sum of products term */
        for ( i=0; i<nvars; i++ ) tterm[i] = 'X';
        while ( outline[k] != '+' ) {
            /* build truth table term */
            if ( outline[k] == '\'' ) {
                vdig = '0';
                k++;
            } else vdig = '1';

            for ( i=0; i<nvars; i++ ) {
                if ( outline[k] == vname[i] ) {
                    tterm[i] = vdig;
                }
            }
            k++;
        }
        /* set up minterms */
        ntrmv(0,tterm,0);
        k++;
    }
```

```
    return;
}
quine()
{
register i, j;
/* initialize prime implicant count */
for ( j=0; j<nwords; j++ ) {
    impcnt[j] = 0;
    impext[j] = 0;
}
for ( j=0; j<MIMPS; j++ ) {
    essprm[j] = 0;
}
/* set up pairings list */
for ( i=0; i<nterms; i++ ) {
    if (gbit((minterm[i/WORD]|noterm[i/WORD]),i%WORD) == 1 ) {
        plist[pptr].nocare = 0;
        plist[pptr].match = 0;
        plist[pptr].mom = 1;
        plist[pptr].dad = 0;
        plist[pptr++].term = i;
            if ( pptr >= MTERMS ) {
                fprintf(stderr,"Too many minterms ( > %d )\n",MTERMS);
                fprintf(stderr,"Process aborted\n");
                i = nterms;
                pptr = 0;          /* nullify process */
            }
    }
}
/* process pairings */
pairup(0,pptr);
return;
}
pairup(first,last)
int first, last;         /* pointers to first term and last term ( + 1 ) */
                            /* of candidate terms at one level of Q-M reduc-*/
                            /* tion.                                         */
{
int match = 0;                    /* indicates a pairing was found             */
int submatch = 0;  /* pairing found on one pass                 */
int diff, diffx = 0; /* nocare term variables                    */
int fterm, dterm;    /* first term in loop parameters             */
register next;               /* pointer to next available plist location  */
int jstart, second;  /* pointers within the level                 */
int i, j2;
register j, k;
```

```
jstart = first;
second = first;
next = last;                  /* initialize loop controls                    */
j = jstart;
j2 = jstart;
while ( jstart< last-1 ) {
   while ( j2 < (last-1) ) {
         for ( k=second+1; k<last; k++ ) {
            /* At this point, the full series of Quine-McCluskey 'tests' */
            /* are made to see if a pairing can be made.               */
            if ((plist[j].nocare == plist[k].nocare ) &&
                    (bitcount(nvars,plist[j].term)
                    == (bitcount(nvars,plist[k].term)-1)) &&
                    (bitcount(nvars,diff=(plist[k].term^plist[j].term))==1) ) {
                 if ((diffx==0)||(((((plist[j].term-fterm)%dterm) == 0) &&
                    (diff==diffx))){

                    /* A pairing has been made. Record the pair at the */
                    /* next level.                                     */
                    match = 1;
                    submatch = 1;
                    if ( diffx == 0 ) {
                            dterm = plist[k].term-plist[j].term;
                            fterm = plist[j].term;
                    }
                    plist[j].match = 1;
                    plist[k].match = 1;
                    plist[next].match = 0;
                    plist[next].nocare = plist[j].nocare|diff;
                    plist[next].term = plist[j].term;
                    plist[next].mom = j;
                    plist[next++].dad = k;
                    second = k;
                    diffx = diff;
                    j = ++k;
                 }
            }
         }

         /* A series of tests is made to limit the number of  */
         /* possible pairings (without forgetting any), in    */
         /* order to accomplish the tabulation recursively.   */
         if ( submatch == 1 ) {
            second += 2;
            j = second;
            submatch = 0;
```

```
            } else {
               j = ++j;
               j2 = j;
               second = j;
            }
      }
   if ( match == 1 ) {
            /* go to the next level of tabulation */
            pairup(last,next);
            j2 = plist[last].mom;
            j = j2;
            second = plist[last].dad;
            next = last;
            match = 0;
            diffx = 0;
      } else {
         jstart++;
         second = jstart;
         j = jstart;
      }
}
/* process the candidate prime implicant */
primes(first,last);
}
bitcount(len,term)
int len;   /* length of string to be counted */
register term;  /* string to be counted */
{
register i;
register count = 0;
for ( i=0; i< len; i++ ) count+=(term>>i)&1;
return(count);
}
primes(first,last)
int first, last;
{
register i, j;
int flag;
int rep;
int match;
/* output prime implicants */
for ( j=first; j<last; j++ ) {
   if ( plist[j].match == 0 ) {
            flag = 0;
            for ( i=0; i<imps; i++ ) {
               /* test to see if candidate is a subset of a larger prime imp */
```

```c
                if (bitcount(nvars,plist[j].nocare) <= bitcount(nvars,pricare[i]) ){
                   if ((((plist[j].nocare|pricare[i]) == (pricare[i])) &&
                        (((~pricare[i])&priterm[i])==((~pricare[i])&plist[j].term))&&
                        ((plist[j].term|priterm[i]|pricare[i]) ==
                        (priterm[i]|pricare[i]))) {
                        flag = 1;
                    }
               /* test to see if candidate will replace a smaller subset */
                } else if
                        (bitcount(nvars,plist[j].nocare)>bitcount(nvars,pricare[i])) {
                        if ((((pricare[i]|plist[j].nocare)==(plist[j].nocare)) &&
                            (((~plist[j].nocare)&plist[j].term) ==
                            ((~plist[j].nocare)&priterm[i])) &&
                            ((priterm[i]|plist[j].term|plist[j].nocare) ==
                            (plist[j].term|plist[j].nocare))) {
                            flag = 2;
                            rep = i;
                        }
                   }
              }
          /* Add a prime implicant to list--no complications */
          if (flag == 0) {
             primepath(j,imps);
             priterm[imps] = plist[j].term;
             pricare[imps] = plist[j].nocare;
             imps++;
             if ( imps >= MIMPS ) {
                    fprintf(stderr,"Warning, overflow of prime implicant chart\n");
                    imps--; /* for protection */
             }

          /* Preform the replacement of a prime implicat with a larger one */
          } else if (flag == 2) {
             primepath(j,rep);
             priterm[rep] = plist[j].term;
             pricare[rep] = plist[j].nocare;
          }
     }
}
return;
}
primepath(j,imp)
register j;            /* start node in Quine-McCluskey */
register imp;                  /* entry in implicant table */
{
if ( j < pptr ) {
```

```
      /* arrival back at original terms */
      impchart[plist[j].term/WORD][imp] |= (1<<(plist[j].term%WORD));
   } else {
      primepath(plist[j].mom,imp);
      primepath(plist[j].dad,imp);
   }
   return;
}
pchart()
{
register i, j, k;
int uncov;
int temp;
char echar;
/* determine coverage of minterms */
for ( i=0; i<nwords; i++ ) {
   for ( j=0; j<imps; j++ ) {
         impcnt[i] |= impchart[i][j];
   }
}
/* determine multiple coverage of minterms */
for ( i=0; i<nterms; i++ ) {
   temp = 0;
   for ( j=0; j<imps; j++ ) {
         temp += (gbit(impchart[i/WORD][j],i%WORD));
   }
   if ( temp >= 2 ) impext[i/WORD] |= (1<<(i%WORD));
}
/* exclude not-care terms from consideration */
for ( i=0; i<nwords; i++ ) {
   /* eliminate not-care cases */
   impcnt[i] &= ~noterm[i];
   impext[i] &= ~noterm[i];
   /* check for prime implicants */
   for ( j=0; j<WORD; j++ ) {
         if ( (gbit(impcnt[i],j) == 1) && (gbit(impext[i],j)==0) ) {
            k = 0;
            while ( gbit(impchart[i][k],j) == 0 ) k++;
            essprm[k] = 1;
         }
   }
}
/* Determine coverage of essential prime implicants and */
/* print out prime implicants */
fprintf(stdout,"\nPrime implicants ( * indicates essential prime implicant )");
for ( i=0; i<imps; i++ ) {
```

```
      if ( essprm[i] == 1 ) {
              echar = '*';
              for ( j=0; j<nwords; j++ ) {
                 impcnt[j] &= ~impchart[j][i];
              }
      } else echar = ' ';
      fprintf(stdout,"\n%c ",echar);
      for ( j=0; j<nvars; j++ ) {
              if ((((pricare[i]>>(nvars-1-j))&1) == 0) {
                 fprintf(stdout,"%c",vname[j]);
                 if ((((priterm[i]>>(nvars-1-j))&1) == 0) {
                         fprintf(stdout,"'");
                 }
              }
      }
      fprintf(stdout,"\t: ");
      for ( j=0; j<nterms; j++ ) {
              if ( gbit(impchart[j/WORD][i],j%WORD) == 1 )
                 fprintf(stdout,"%d,",j);
      }
}
uncov = 0;
for ( i=0; i<nwords; i++ ) {
   uncov += bitcount(WORD,impcnt[i]);
}
return(uncov);
}
reduction(uncov)
int uncov;
{
register i,j;
int nonemps;                /* number of non-essential terms */
int terms, lits;    /* minimization factors */
int nons[MIMPS]; /* index into impchart of non-essential impl.*/
int termlist[QTERMS];       /* temporary location of covered term count */
int fail;         /* new candidate flag                             */
long limit, li;             /* power set bits                                 */
char oper;                  /* sum of products separator                      */
/* current best coverage candidate */
struct current {
        int terms;
        int lits;
        int list[MIMPS];
        } curbest;
if ( uncov == 0 ) {
   fprintf(stdout,"\n\nminimal expression is unique\n");
```

```c
      } else {
         fprintf(stdout,"\n\nno unique minimal expression\n");
         /* set up non-essential implicant list */
         j = 0;
         for ( i=0; i<imps; i++ )
                 if (essprm[i] == 0) nons[j++] = i;
         nonemps = j;
/* insure no overflow of cyclical prime implicant array */
         if ( nonemps > 2*WORD ) {
                 fprintf(stderr,"Warning! Only %d prime implicants can be",2*WORD);
                 fprintf(stderr,"considered for coverage\n of all terms (in addition");
                 fprintf(stderr,"to essential primes). %d implicants not checked\n",
                    nonemps-(2*WORD));
                 nonemps = 2*WORD;
         }
         if ( nonemps > WORD ) {
                 fprintf(stdout,"Warning! Large number of cyclical prime implicants\n");
                 fprintf(stdout,"Computation will take awhile\n");
         }
         /* candidate coverage is determined by generation of the power set */
         /* calculate power set */
         limit = 1;
         for ( i=0; i<nonemps; i++ ) limit *= 2;
         /* set up current best expression list */
         curbest.terms = BIG;
         curbest.lits = BIG;
         curbest.list[0] = -1;
         /* try each case */
         for ( li=1L; li<limit; li++ ) {
                 terms = bitcount(2*WORD,li);
                 if ( terms <= curbest.terms ) {
                    /* reset count */
                    lits = 0;
                    /* reset uncovered term list */
                    for ( i=0; i<nwords; i++ )
                            termlist[i] = impcnt[i];
                    for ( i=0; i<nonemps; i++ ) {
                            if (((li>>i)&1L) == 1L) {
                               for ( j=0; j<nterms; j++ ) {
                                       if (gbit(impchart[j/WORD][nons[i]],j%WORD)==1) {
                                          termlist[j/WORD] &= ~(1<<(j%WORD));
                                       }
                               }
                            }
                            lits += (nvars - bitcount(nvars,pricare[nons[i]]));
                    }
```

```
                fail = 0;
                for ( i=0; i<nwords; i++ ) {
                        if ( termlist[i]!=0 ) fail = 1;
                }
                if ((fail==0) && ((terms<curbest.terms) || (lits<curbest.lits))) {
                        /* we have a new candidate */
                        curbest.terms = terms;
                        curbest.lits = lits;
                        j = 0;
                        for ( i=0; i<nonemps; i++ ) {
                           if (((li>>i)&1L)==1L) {
                                    curbest.list[j++] = nons[i];
                           }
                        }
                        curbest.list[j] = -1;
                }
            }
    }

    j = 0;
    while ( curbest.list[j] >= 0 ) {
            essprm[curbest.list[j]] = 1;
            j++;
    }
}
/* print out minimal expression */
fprintf(stdout,"\nminimal expression:\n\n    %c(",func);
for ( i=0; i<nvars; i++ ) {
   fprintf(stdout,"%c",vname[i]);
   if ( i < nvars - 1 ) fprintf(stdout,",");
}
fprintf(stdout,")");
oper = '=';
for ( i=0; i<imps; i++ ) {
   if ( essprm[i] == 1 ) {
           fprintf(stdout," %c ",oper);
           for ( j=0; j<nvars; j++ ) {
              if (((pricare[i]>>(nvars-1-j))&1) == 0) {
                 fprintf(stdout,"%c",vname[j]);
                 if (((priterm[i]>>(nvars-1-j))&1) == 0) {
                         fprintf(stdout,"'");
                 }
              }
           }
           oper = '+';
   }
```

```
}
fprintf(stdout,"\n\n");
return;
}
```

The above code is a recursive implementation of the Quine-McCluskey Method which also includes case for don't care conditions.

## Code for Quine-McCluskey Method in C++

### Variant-I

```cpp
#include <iostream>
#include <vector>
#include <string>
#include <stdlib.h>
using namespace std;

int MIN_BITS = 4;              //minimum bits to print
vector<unsigned> input_values;
bool show_mid = false;         //show middle process

struct B_number{
        unsigned number;
        unsigned dashes;
        bool used;
};

vector<vector<B_number> > table;       //original table
vector<vector<B_number> > p_group;     //mid process table
vector<vector<B_number> > final_group; //final table
vector<B_number> printed_numbers; //avoid printing the same final numbers
//-----------------------------------------------------------
unsigned count_1s(unsigned number); //count the number of 1s in a number
void print_binary(unsigned number);//print the number in binary
void create_table();           //create original table sorted by the number of 1s
void print_table();            //print the table
B_number init_B_number(unsigned n,int d, bool u);//initialize a B_number
void create_p_group();         //create mid process table
void print_p_group();          //print it
void print_p_binary(unsigned n, unsigned d);//print the mid table (with -'s)
void create_final_group();     //create final table
void print_final_group();      //print final table with -'s and unused terms
bool is_printed(B_number n);   //dont print terms that were already printed
void init();                   //start the table making and printing
void getinput();               //get input from user
unsigned count_bits(unsigned n);    //min bits to represent a number
//-----------------------------------------------------------
```

```
int main(int argc, char *argv[])
{
        /* allow command line calling with arguments -m -b X
           where X is a number. order or -m and -b X does not
           matter*/
        cout<<"\nQMCS - Quine McCluskey Simplifier\n";
        if(argc >= 2)
        {
                string arg = argv[1];
                if(arg.find("-m") != -1) {
                        show_mid = true;
                        if(argc>=3) {
                                arg = argv[2];
                                if(arg.find("-b") != -1)
                                        MIN_BITS = atoi(argv[3]);
                        }
                }
                else if(arg.find("-h") != -1) {
                        cout<<"-b X\tminimum bits should be X.\n"
                           <<"-m  \tshow mid process computation.\n"
                           <<"-h  \tshow this.\n";
                        return 0;
                }
                else {
                        if(arg.find("-b") != -1  && argc >=3)
                                MIN_BITS = atoi(argv[2]);

                        if(argc >=4) {
                                arg = argv[3];
                                if(arg.find("-m") != -1)
                                        show_mid = true;
                        }
                        else
                        {
                                cout<<"Invalid argument\n"
                                   <<"-b X\tminimum bits should be X.\n"
                                   <<"-m  \tshow mid process computation.\n"
                                   <<"-h  \tshow this.\n";
                                return 0;
                        }
                }
        }
        getinput();
        init();
```

```
                return 0;
}
/* counts 1s by getting the LSB (%2) and then shifting until 0 */
unsigned count_1s(unsigned number) {
        short bit =0;
        int count = 0;
        while(number>0)      {
                bit = number%2;
                number>>=1;
                if(bit) {
                        count++;
                }
        }
        return count;
}
/*get LSB, arrange it in array, the print array in reverse order so MSB is on
the left */
void print_binary(unsigned number) {
        unsigned bits[MIN_BITS];
        int count = 0;
        while(number>0||count<MIN_BITS) {
                bits[count] = number%2;
                number>>= 1;
                count++;
        }
        for(int i=count-1;i>=0;i--)
                cout<<bits[i];
}
/*creating first table: append current number to the array located in
table[number of 1s f this number]*/
void create_table() {
        short tmp;
        B_number temp_num;
        for(int i=0;i<input_values.size();i++) {
                tmp = count_1s(input_values[i]);
                if(tmp+1>table.size())
                        table.resize(tmp+1);

                temp_num = init_B_number(input_values[i],0,false);
                table[tmp].push_back(temp_num);
        }
}
void print_table() {
        cout<<endl<<"COMPUTING:"<<endl;
        for(int i=0;i<table.size();i++) {
                cout<<i;
```

```cpp
            for(int j=0;j<table[i].size();j++) {
                cout<<"\tm"<<table[i][j].number<<"\t";
                print_binary(table[i][j].number);
                cout<<endl;
            }
            cout<<"\n-----------------------------------"<<endl;
        }
    }
    /* initialize a B_number variable - like a constructor */
    B_number init_B_number(unsigned  n,int d, bool u) {
        B_number num;
        num.number = n;
        num.dashes = d;
        num.used = u;
        return num;
    }
    /*like the original table, but the paring of numbers from the original table-
    dashes are represented by a 1. for example original A=0011 B=1011, new number
    is -011 which is represented as C.number=A&B=0011,C.dashes=A^B=1000*/
    void create_p_group() {
        short tmp;
        B_number temp_num;
        unsigned temp_number, temp_dashes;
        for(int i=0;i<table.size()-1;i++) {
            for(int j=0;j<table[i].size();j++) {
                for(int k=0;k<table[i+1].size();k++) {
                    temp_number = table[i][j].number & table[i+1][k].number;
                    temp_dashes = table[i][j].number ^ table[i+1][k].number;
                        if(count_1s(temp_dashes)==1) {
                        table[i][j].used = true;
                        table[i+1][k].used = true;

                        tmp = count_1s(temp_number);

                        if(tmp+1>p_group.size())
                            p_group.resize(tmp+1);

                        temp_num    =    init_B_number(temp_number, temp_dashes, false);
                        p_group[tmp].push_back(temp_num);
                    }
                }
            }
        }
    }
```

```cpp
void print_p_group() {
	cout<<endl<<"MID PROCESS COMPUTATION:"<<endl;
	for(int i=0;i<p_group.size();i++) {
		cout<<i;
		for(int j=0;j<p_group[i].size();j++) {
			cout<<"\t\t";
			print_p_binary(p_group[i][j].number,p_group[i][j].dashes);
			cout<<endl;
		}
		cout<<"\n------------------------------------"<<endl;
	}
}
/*print a number such as -001; this allocates bits in an array dash=2 then
prints reverse array */
void print_p_binary(unsigned n, unsigned d) {
	unsigned bits[MIN_BITS];
	int count = 0;
	while(n>0||count<MIN_BITS) {
		if(!(d%2))
			bits[count] = n%2;
		else
			bits[count] = 2;
		n >>= 1;
		d >>= 1;
		count++;
	}
	for(int i=count-1;i>=0;i--) {
		if(bits[i]!=2)
			cout<<bits[i];
		else
			cout<<"-";
	}
}
/*creates final table. works like p_group(). example; in p_group you have:
A=-001 B=-011 -> C= -0-1 which will be represented as
C.number=A&B=0001&0011=0001,                                          and
C.dashes=A^B^A.dashes=0001^0011^1000=1010.
Computation is done only when A.dashes = b.dashes*/
void create_final_group() {
	short tmp;
	B_number temp_num;
	unsigned temp_number, temp_dashes;
	for(int i=0;i<p_group.size()-1;i++) {
		for(int j=0;j<p_group[i].size();j++) {
			for(int k=0;k<p_group[i+1].size();k++) {
```

```cpp
                            if(p_group[i][j].dashes == p_group[i+1][k].dashes) {
                    temp_number = p_group[i][j].number & p_group[i+1][k].number;
                    temp_dashes = p_group[i][j].number ^ p_group[i+1][k].number;
                    if(count_1s(temp_dashes)==1)
                            {
                                            temp_dashes^= p_group[i][j].dashes;

                                            p_group[i][j].used = true;
                                            p_group[i+1][k].used = true;

                                            tmp = count_1s(temp_number);

                                            if(tmp+1>final_group.size())
                                                    final_group.resize(tmp+1);

                                             temp_num = init_B_number(temp_number, temp_dashes, true);
                                            final_group[tmp].push_back(temp_num);
                                }
                            }
                        }
                }
            }
    }
}
/*print all the values from the final table, except for duplicates.
 print all the unused numbers from original table and mid process table*/
void print_final_group() {
        cout<<endl<<"FINAL:\n-------------------------------------"<<endl;
        int i,j;
        for(i=0;i<final_group.size();i++) {
                for(j=0;j<final_group[i].size();j++) {
                        if(!is_printed(final_group[i][j])) {

        print_p_binary(final_group[i][j].number,final_group[i][j].dashes);
                                cout<<endl;
                                printed_numbers.push_back(final_group[i][j]);
                        }
                }
        }
        for(i=0;i<p_group.size();i++) {
                for(j=0;j<p_group[i].size();j++) {
                        if(!p_group[i][j].used) {
                                print_p_binary(p_group[i][j].number,p_group[i][j].dashes);
                                cout<<endl;
                        }
                }
```

```cpp
            }
        for(i=0;i<table.size();i++) {
                for(j=0;j<table[i].size();j++) {
                        if(!table[i][j].used) {
                                print_p_binary(table[i][j].number,table[i][j].dashes);
                                cout<<endl;
                        }
                }
        }
        cout<<"------------------------------------"<<endl;
}
/*used to avoid printing duplicates that can exist in the final table*/
bool is_printed(B_number n) {
        for(int i=0;i<printed_numbers.size();i++)
                if(n.number==printed_numbers[i].number && n.dashes == printed_numbers[i].dashes)
                        return true;
        return false;
}
/*initialize all table*/
void init() {
        table.resize(1);
        p_group.resize(1);
        final_group.resize(1);
        create_table();
        print_table();
        create_p_group();
        if(show_mid)
                print_p_group();
        create_final_group();
        print_final_group();
}
void getinput() {
        unsigned in;
        int num_bits=0;
        cout<<"\nInput value followed by ENTER[^D ends input]\n> ";
        while(cin>>in) {
                input_values.push_back(in);
                num_bits = count_bits(in);
                if(num_bits>MIN_BITS)
                        MIN_BITS = num_bits;
                cout<<"> ";
        }
}
/*return min number of bits a number is represented by. used for best output*/
unsigned count_bits(unsigned n) {
```

```
        short bit =0;
        int count = 0;
        while(n>0) {
                bit = n%2;
                n>>=1;
                count++;
        }
        return count;
}
```

The following code implements the Quine-McCluskey method but the implementation is bit sloppy and a more optimized version of this code is given in variant-II.

**Variant-II**

```
#include <iostream>
#include <iomanip>
#include <sstream>
#include <string>
#include <vector>
#include <algorithm>
using namespace std;

int count1s(size_t x) {
        int o = 0;
        while (x) {
                o += x % 2;
                x >>= 1;
        }
        return o;
}
int vars;
struct Implicant {
        int implicant;
        string minterms;
        vector<int> mints;
        int mask;
        string bits;
        int ones;
        bool used;
        Implicant(int i = 0, vector<int> min = vector<int>(), string t = "", int m = 0, bool u = false)
                : implicant(i), mask(m), ones(0), used(u)
        {
                if (t == "") {
                        stringstream ss;
                        ss << 'm' << i;
```

```
                                minterms = ss.str();
                        }
                        else minterms = t;
                        if (min.empty()) mints.push_back(i);
                        else mints = min;
                        int bit = 1 << vars;
                        while (bit >>= 1)
                                if (m & bit) bits += '-';
                                else if (i & bit) { bits += '1'; ++ones; }
                                else bits += '0';
                }
                bool operator<(const Implicant& b) const { return ones < b.ones; }
                vector<int> cat(const Implicant &b) {
                        vector<int> v = mints;
                        v.insert(v.end(), b.mints.begin(), b.mints.end());
                        return v;
                }
                friend ostream &operator<<(ostream &out, const Implicant &im);
        };
bool pr = true;
bool fin = true;
ostream &operator<<(ostream &out, const Implicant &im) {
        int bit = 1 << vars, lit = 0;
        ostringstream ss;
        while (bit >>= 1) {
                if (!(im.mask & bit))
                        ss << char(lit + 'A') << (im.implicant & bit ? ' ' : '\'');
                ++lit;
        }
        if (fin) out << right << setw(16);
        out << ss.str();
        if (pr) out << '\t' << setw(16) << left << im.minterms << ' ' << im.bits << '\t' << im.ones;
        return out;
}
void printTab(const vector<Implicant> &imp) {
        for (size_t i = 0; i < imp.size(); ++i)
                cout << imp[i] << endl;
        cout << "-------------------------------------------------\n";
}
void mul(vector<size_t> &a, const vector<size_t> &b) {
        vector<size_t> v;
        for (size_t i = 0; i < a.size(); ++i)
                for (size_t j = 0; j < b.size(); ++j)
                        v.push_back(a[i] | b[j]);
        sort(v.begin(), v.end());
```

```
            v.erase( unique( v.begin(), v.end() ), v.end() );
            for (size_t i = 0; i < v.size() - 1; ++i)
                    for (size_t j = v.size() - 1; j > i ; --j) {
                            size_t z = v[i] & v[j];
                            if ((z & v[i]) == v[i])
                                    v.erase(v.begin() + j);
                            else if ((z & v[j]) == v[j]) {
                                    size_t t = v[i];
                                    v[i] = v[j];
                                    v[j] = t;
                                    v.erase(v.begin() + j);
                                    j = v.size();
                            }
                    }
            a = v;
    }
    int main() {
            cin >> vars;
            int combs = 1 << vars;
            //bool *minterms = (bool *)calloc(combs, sizeof(bool));
            vector<int> minterms;
            vector<Implicant> implicants;
            for (int mint; cin >> mint; ) {
                    implicants.push_back(mint);
                    minterms.push_back(mint);
            }
            if (!minterms.size()) { cout << "\n\tF = 0\n"; return 0; }
            sort(minterms.begin(), minterms.end());
            minterms.erase( unique( minterms.begin(), minterms.end() ), minterms.end() );
            if (!cin.eof() && cin.fail()) {  // don't cares
                    cin.clear();
                    while ('d' != cin.get()) ;
                    for (int mint; cin >> mint; )
                            implicants.push_back(mint);
            }
            sort(implicants.begin(), implicants.end());
            printTab(implicants);
            vector<Implicant> aux;
            vector<Implicant> primes;
            while (implicants.size() > 1) {
                    for (size_t i = 0; i < implicants.size() - 1; ++i)
                            for (size_t j = implicants.size() - 1; j > i ; --j)
                                    if (implicants[j].bits == implicants[i].bits)
                                            implicants.erase(implicants.begin() + j);
                    aux.clear();
                    for (size_t i = 0; i < implicants.size() - 1; ++i)
```

```
                    for (size_t j = i + 1; j < implicants.size(); ++j)
                        if (implicants[j].ones == implicants[i].ones + 1 &&
                            implicants[j].mask == implicants[i].mask &&
                            count1s(implicants[i].implicant ^
implicants[j].implicant) == 1) {
                                implicants[i].used = true;
                                implicants[j].used = true;
                                aux.push_back(
                                    Implicant(implicants[i].implicant,
                                    implicants[i].cat(implicants[j]),
                                    implicants[i].minterms + ',' +
implicants[j].minterms,
                                    (implicants[i].implicant ^
implicants[j].implicant) | implicants[i].mask)
                                );
                        }
            for (size_t i = 0; i < implicants.size(); ++i)
                if (!implicants[i].used) primes.push_back(implicants[i]);
            implicants = aux;
            sort(implicants.begin(), implicants.end());
            printTab(implicants);
    }
    for (size_t i = 0; i < implicants.size(); ++i)
        primes.push_back(implicants[i]);
    if (primes.back().mask == combs - 1)
        { cout << "\n\tF = 1\n"; return 0; }
    pr = false;
    bool table[primes.size()][minterms.size()];
    for (size_t i = 0; i < primes.size(); ++i)
        for (size_t k = 0; k < minterms.size(); ++k)
            table[i][k] = false;
    for (size_t i = 0; i < primes.size(); ++i)
        for (size_t j = 0; j < primes[i].mints.size(); ++j)
            for (size_t k = 0; k < minterms.size(); ++k)
                if (primes[i].mints[j] == minterms[k])
                    table[i][k] = true;
    for (int k = 0; k < 18; ++k) cout << " ";
    for (size_t k = 0; k < minterms.size(); ++k)
        cout << right << setw(2) << minterms[k] << ' ';
    cout << endl;
    for (int k = 0; k < 18; ++k) cout << " ";
    for (size_t k = 0; k < minterms.size(); ++k)
        cout << "---";
    cout << endl;
    for (size_t i = 0; i < primes.size(); ++i) {
        cout << primes[i] << " |";
```

```cpp
            for (size_t k = 0; k < minterms.size(); ++k)
                    cout << (table[i][k] ? " X " : "   ");
            cout << endl;
    }
    vector<size_t> M0, M1;
    for (size_t i = 0; i < primes.size(); ++i)
            if (table[i][0]) M0.push_back(1 << i);
    for (size_t k = 1; k < minterms.size(); ++k) {
            M1.clear();
            for (size_t i = 0; i < primes.size(); ++i)
                    if (table[i][k]) M1.push_back(1 << i);
            mul(M0, M1);
    }
    int min = count1s(M0[0]);
    size_t ind = 0;
    // for (size_t i = 0; i < M0.size(); ++i) cout << M0[i] << ','; cout << endl;
    for (size_t i = 1; i < M0.size(); ++i)
            if (min > count1s(M0[i])) {
                    min = count1s(M0[i]);
                    ind = i;
            }
    fin = false;
    bool f;
    cout << "-------------------------------------------------------\n";
    for (size_t j = 0; j < M0.size(); ++j) {
            cout << "\tF = ";
            f = false;
            for (size_t i = 0; i < primes.size(); ++i)
                    if (M0[j] & (1 << i)) {
                            if (f) cout << " + ";
                            f = true;
                            cout << primes[i];
                    }
            cout << endl;
    }
    cout << "-------------------------------------------------------\n";
    // minimal solution
    cout << "\n\tF = ";
    f = false;
    for (size_t i = 0; i < primes.size(); ++i)
            if (M0[ind] & (1 << i)) {
                    if (f) cout << " + ";
                    f = true;
                    cout << primes[i];
            }
    cout << endl;
```

}

This is a more optimized code as compared to the previous one.

## V. COMPLEXITY

Although more practical than Karnaugh mapping when dealing with more than four variables, the Quine–McCluskey algorithm also has a limited range of use since the problem it solves is NP-hard: the runtime of the Quine–McCluskey algorithm grows exponentially with the number of variables. It can be shown that for a function of n variables the upper bound on the number of prime implicants is 3n/n. If n = 32 there may be over 6.5 * 1015 prime implicants. Functions with a large number of variables have to be minimized with potentially non-optimal heuristic methods, of which the Espresso heuristic logic minimizer is the de facto standard.

## VI. CONCLUSION

In this paper we have listed the codes for the implementation of Quine-McCluskey method using the computer languages C and C++. Readers well versed in any of these languages would be at ease to follow with the computer code. In preparing these codes, one primary observation we had was that the number of lines of code in C++ was about 100 lines less than what could be achieved in C. This provides us an insight about the inherent advantage we get in the object oriented design paradigm as compared to the procedural languages. Understanding the concept and theory behind the Quine-McCluskey method is the key to writing good and optimized codes in any of the computer languages.